\pdfoutput=1

\documentclass[journal]{IEEEtran}
\usepackage{cite}
\ifCLASSINFOpdf
\else
\fi
\usepackage[cmex10]{amsmath,mathtools}
\usepackage{amsfonts,amssymb}

\usepackage{array}
\usepackage{amsfonts}
\usepackage{amsmath,amsthm}
\usepackage{tikz}
\usetikzlibrary{shapes.geometric, arrows}

\tikzstyle{rect} = [rectangle, minimum width=2cm, minimum height=1cm, text centered, draw=black, fill=blue!3]
\tikzstyle{arrow} = [thick,->,>=stealth]

\usepackage{algorithm}
\usepackage{algcompatible} % With the help of this we can use Statex do introduce unnumbered lines.
\usepackage{algpseudocode}

\usepackage{setspace}

\usepackage[latin1]{inputenc} % So you can write swedish letters.

\usepackage{graphicx}
\usepackage{epstopdf}
\usepackage[process=auto]{pstool}
\usepackage{psfrag}

\usepackage{wrapfig}

\usepackage{relsize}
\usepackage{transparent}

\newcommand{\algorithmfootnote}[2][\footnotesize]{%
  \let\old@algocf@finish\@algocf@finish% Store algorithm finish macro
  \def\@algocf@finish{\old@algocf@finish% Update finish macro to insert "footnote"
    \leavevmode\rlap{\begin{minipage}{\linewidth}
    #1#2
    \end{minipage}}%
  }%
}

\begin{document}
%
% paper title
% can use linebreaks \\ within to get better formatting as desired
% Do not put math or special symbols in the title.
%\title{Map-aided Dead-reckoning using only Measurements of Speed}
\title{Map-aided Dead-reckoning --- A Study on \\Locational Privacy in Insurance Telematics}
%\title{A Study on Locational Privacy in Insurance Telematics Programs collecting Speed Measurements}
%
%
% author names and IEEE memberships
% note positions of commas and nonbreaking spaces ( ~ ) LaTeX will not break
% a structure at a ~ so this keeps an author's name from being broken across
% two lines.
% use \thanks{} to gain access to the first footnote area
% a separate \thanks must be used for each paragraph as LaTeX2e's \thanks
% was not built to handle multiple paragraphs
%

\pagenumbering{gobble} % Removes page numbers.
\author{Johan Wahlstr\"om, Isaac Skog,~\IEEEmembership{Member,~IEEE}, João G. P. Rodrigues,~\IEEEmembership{Student Member,~IEEE}, Peter H\"andel,~\IEEEmembership{Senior Member,~IEEE}, and Ana Aguiar,~\IEEEmembership{Member,~IEEE}
\thanks{
J. Wahlstr\"{o}m, I. Skog, and P. H\"{a}ndel are with the ACCESS Linnaeus Center, Dept. of Signal Processing, KTH Royal Institute of Technology, Stockholm, Sweden. (e-mail:  \{jwahlst, skog, ph\}@kth.se).

J. G. P. Rodrigues and A. Aguiar are with the Instituto de Telecomunicações, Departamento de Engenharia Eletrotécnica e de Computadores, Faculdade de Engenharia da Universidade do Porto, 4200-465 Porto, Portugal. (e-mail: \{joao.g.p.rodrigues, anaa\}@fe.up.pt)
}
%\thanks{Manuscript received April --, ----; revised December --, ----.}
}

% note the % following the last \IEEEmembership and also \thanks -
% these prevent an unwanted space from occurring between the last author name
% and the end of the author line. i.e., if you had this:
%
% \author{....lastname \thanks{...} \thanks{...} }
%                     ^------------^------------^----Do not want these spaces!
%
% a space would be appended to the last name and could cause every name on that
% line to be shifted left slightly. This is one of those "LaTeX things". For
% instance, "\textbf{A} \textbf{B}" will typeset as "A B" not "AB". To get
% "AB" then you have to do: "\textbf{A}\textbf{B}"
% \thanks is no different in this regard, so shield the last } of each \thanks
% that ends a line with a % and do not let a space in before the next \thanks.
% Spaces after \IEEEmembership other than the last one are OK (and needed) as
% you are supposed to have spaces between the names. For what it is worth,
% this is a minor point as most people would not even notice if the said evil
% space somehow managed to creep in.

% The paper headers
%\markboth{IEEE TRANSACTIONS ON INTELLIGENT TRANSPORTATION SYSTEMS, ~Vol.~--, No.~-, December~----}%
%\markboth{IEEE, ~Vol.~--, No.~-, December~----}%
\markboth{}%
{Shell \MakeLowercase{\textit{et al.}}: Bare Demo of IEEEtran.cls for Journals}
% The only time the second header will appear is for the odd numbered pages
% after the title page when using the twoside option.
%
% *** Note that you probably will NOT want to include the author's ***
% *** name in the headers of peer review papers.                   ***
% You can use \ifCLASSOPTIONpeerreview for conditional compilation here if
% you desire.

% If you want to put a publisher's ID mark on the page you can do it like
% this:
%\IEEEpubid{0000--0000/00\$00.00~\copyright~2012 IEEE}
% Remember, if you use this you must call \IEEEpubidadjcol in the second
% column for its text to clear the IEEEpubid mark.

% use for special paper notices
%\IEEEspecialpapernotice{(Invited Paper)}

% make the title area
\maketitle

% As a general rule, do not put math, special symbols or citations
% in the abstract or keywords.

\vspace*{-8mm}

\begin{abstract}
We present a particle-based framework for estimating the position of a vehicle using map information and measurements of speed. Two measurement functions are considered. The first is based on the assumption that the lateral force on the vehicle does not exceed critical limits derived from physical constraints. The second is based on the assumption that the driver approaches a target speed derived from the speed limits along the upcoming trajectory. Performance evaluations of the proposed method indicate that end destinations often can be estimated with an accuracy in the order of $100\,[m]$. These results expose the sensitivity and commercial value of data collected in many of today's insurance telematics programs, and thereby have privacy implications for millions of policyholders. We end by discussing the strengths and weaknesses of different methods for anonymization and privacy preservation in telematics programs.
\end{abstract}

% Note that keywords are not normally used for peerreview papers.

% For peer review papers, you can put extra information on the cover
% page as needed:
% \ifCLASSOPTIONpeerreview
% \begin{center} \bfseries EDICS Category: 3-BBND \end{center}
% \fi
%
% For peerreview papers, this IEEEtran command inserts a page break and
% creates the second title. It will be ignored for other modes.
\IEEEpeerreviewmaketitle

\section{Introduction}

\begin{figure}[t]
\vspace*{-5mm}
\def\svgwidth{0.8\columnwidth}
\hspace*{0mm}
\scalebox{1}{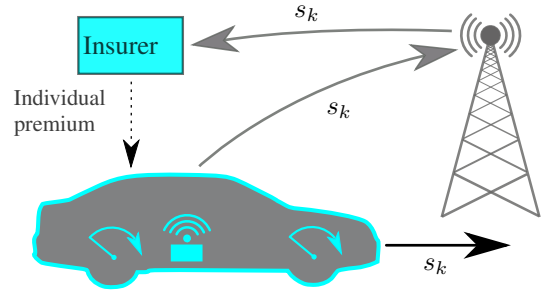}
\vspace*{-1mm}
\caption{Several insurance telematics providers are today collecting measurements of speed, here denoted by $s_k$, from vehicles' wheel speed sensors to adjust premiums and provide feedback to drivers.}
\label{telematics_illustration}
\vspace*{-1mm}
\end{figure}

The emergence of connected vehicles has opened up possibilities for services which combine the power of cloud computing with the expanding capabilities of modern vehicle technology.
%This interdisciplinary approach
Connected vehicles not only enable remote analysis of vehicle condition and driving behavior, but also allow drivers to benefit from a multitude of convenience-related applications such as theft tracking and accident detection. However, as the amounts of generated data and the means of connectivity increases, so do the dangers related to privacy and security \cite{Lawson2015}. Recently, connected vehicles and associated aftermarket devices have been shown to be vulnerable to wireless hacking via a broad range of attack vectors \cite{Pike2015,Checkoway2011,Foster2015}. In the worst case, the hacker will be able to remotely manipulate any of the vehicle's electronic control units (ECUs). Moreover, privacy concerns have been raised about the large databases of sensitive data collected from vehicle-installed telematics units \cite{Duri2004,Rizzo2015,Gao2014,Dewri2013}. As illustrated in Fig. \ref{telematics_illustration}, a telematics insurer uses data on driving behavior to set the premium offered to each policyholder \cite{Handel2014,Wahlstrom2015c,Engelbrecht2015}. Typically, the risk profile of each driver will be based on e.g., the number of speeding or harsh braking events, as detected using measurements from vehicle-installed black boxes or from the vehicle's on-board-diagnostics (OBD) system \cite{Wahlstrom2015b,Engelbrecht2014}. This study examines the locational privacy of such insurance telematics programs.

\subsection{Locational Privacy}

The digital revolution has made violations of locational privacy both cheaper and harder to detect \cite{Blumberg2009}. Temporally isolated digital position inferences were first made possible by the widespread usage of credit cards and access cards. Today, the abundance of e.g., global navigation satellite system (GNSS) receivers in smartphones and vehicles enables continuous streams of position measurements to be sent to central servers \cite{Iqbal2010,Michael2006}. In addition, it is often possible to perform smartphone-based WiFi-localization by exploiting system permissions required by many popular apps \cite{Nguyen2013}. %Some will also argue that consumers are willingly giving up personal information in exchange for services and products. However, even if data is shared with the user's consent, the actual privacy implications of sharing specific data are often obscured.
Although some might argue that consumers are willingly giving up personal information in exchange for services and products, the actual privacy implications of sharing specific data are often obscured. This is especially true when the data itself does not include position measurements, but rather, information which in combination with additional databases can be used for positional inference.

Initially, privacy and security were not major concerns in the insurance telematics industry, and the amount of data collected was primarily determined by restrictions on transmission and storage. However, these topics have received more attention as the industry has matured, and privacy concerns have even been cited as a contributing reason to suspend telematics trials \cite{Troncoso2011,Paefgen2013,Courtney2013,Ohlsson2015}. Today, several telematics providers only collect measurements of speed. When responding to privacy concerns, these companies will refer to the fact that no position data is recorded. However, as shown in \cite{Gao2014}, location-based information can still be extracted from many trips.

\subsection{Contributions and Outline}

In this article, we use map-aided dead-reckoning (DR) to geographically locate a vehicle purely based on its speed profile and digital map information. Since the study utilizes the same type of information that is available for many of the current insurance telematics providers, the results have privacy implications for millions of policyholders. The estimation is carried out using a particle filter, and thereby parallels previous filtering-based approaches where measurements of the vehicle's yaw rate have been available \cite{Hall2001, Svenzen2002, Hedlund2008, Kronander2004}. Though the idea of estimating a vehicle's position using only speed measurements and map information is not new \cite{Gao2014,Dewri2013}, this is the first time that it has been formulated as a filtering problem. In addition to making it possible to motivate the implementation using optimality arguments, our hope is that this also will make the method more accessible, and facilitate extensions based on established methods for particle filtering. Previous algorithms have been very reliant on the matching of vehicle stops in the speed data with intersections or traffic signals in the map. For example, in \cite{Dewri2013}, data where the vehicle made a large number of stops due to traffic congestion or similar was removed altogether since this type of driving was not accounted for in the algorithm. In this study, we avoid these issues by processing the speed measurements one-by-one, rather than in segments delimited by vehicle stops. Moreover, the performance evaluations are made on data sets with significantly longer trip lengths than in previous studies.

Section \ref{preliminaries_section} reviews the state-of-the-art in map-aided dead-reckoning and discusses how the problem is altered depending on what sensor information is available. The employed system model and the filter implementation are detailed in Section \ref{Model_Estimation_Section}. Section \ref{Field_Study_Section} illustrates the performance characteristics of the proposed method by studying the accuracy of estimated end positions and the dependence on a priori knowledge of the initial position. The results indicate that GNSS or OBD measurements of speed often are sufficient to extract a substantial amount personal information on daily activities and location-based behavior. Section \ref{discussion_section} discusses the implications of the presented results and possible ways forward for privacy-aware telematics providers. Finally, the article is concluded in Section \ref{conclusions_section}.

\section{Preliminaries on Map-aided Dead-reckoning}
\label{preliminaries_section}

DR refers to the process of recursively estimating the position of an object by numerically integrating measurements or estimates of velocity. The general concept has been widely used to navigate e.g., aircrafts, ships, submarines, robots, automobiles, and pedestrians \cite{Groves2008}. Obviously, perfect integration requires perfect knowledge of the velocity at all time instances. In practice, this means that stand-alone dead-reckoning systems always will be subject to position errors that accumulate with time. To prevent the position error from growing without bound, additional information is required. This type of information can be provided by, e.g., satellite-based positioning systems \cite{Wahlstrom2015b}, WiFi fingerprinting systems \cite{Chen2002}, magnetic fingerprinting systems \cite{Haverinen2009}, terrain measurements \cite{Bergman97}, or road maps.

\subsection{Map-aided DR with Sensor-based Yaw Information}

DR aided only by information from road maps (and some a priori knowledge of the initial position) have been the topic of several studies within land-vehicle navigation \cite{Gustafsson2012}. Historically, the primary motivation of these studies have been to increase the accuracy and integrity of existing low-cost navigation systems during failures or performance deteriorations of the employed positioning systems (e.g., during GNSS outages). Assuming that measurements of both left and right wheel speeds are available from the controller area network (CAN) bus, estimates of the vehicle's longitudinal velocity and yaw rate can be obtained by means of differential geometry \cite{Hall2001}. These estimates are then used for DR. If only a scalar speed measurement is available at each sampling instance, measurements of the yaw angle or yaw rate can instead be obtained from the steering wheel angle sensor, from a vehicle-mounted gyroscope, or from a magnetometer \cite{Davidson2010}. The error growth of the navigation solution is typically mitigated by comparing the current position and yaw estimates with the corresponding quantities at the nearest map segment \cite{Skog2009,Svenzen2002}. To avoid the logistics of accessing measurements of wheel speed, DR can instead be performed by means of inertial navigation. However, this will increase the number of integration steps, and thereby also the rate of the position error growth. As a result, the position estimates will be very sensitive to e.g., mounting misalignments and sensor biases. In general, the performance of inertial measurement unit (IMU)-based methods can be expected to be inferior to that of odometric methods \cite{Hedlund2008}.

\subsection{Map-aided DR without Sensor-based Yaw Information}

Since the standard parameters available from the vehicle's OBD system do not include any information on the vehicle's yaw rate or yaw angle, it is interesting to see what level of positioning accuracy that can be obtained by using only OBD measurements of speed. (As previously mentioned, the industry of insurance telematics today includes several actors, e.g., Progressive, Geico, and Allstate, that collect speed measurements from their policyholders \cite{Ptolemus2014,Derikx2015}.) Related studies were conducted in \cite{Gao2014} and \cite{Dewri2013}, which estimated the position of a vehicle by matching intersections in the map with detected vehicle stops in the speed measurements. A list of possible map traces was continuously updated by adding new traces after each vehicle stop, and deleting traces deemed infeasible after 1) comparing their lengths with the corresponding distance implied by the speed measurements; 2) comparing the measured speeds with speed limits; and 3) comparing the measured speeds with the maximum speeds as implied by the curvature of the road. The estimation was not statistically motivated and was not formulated within any established estimation framework. In this study, we aim to fill this gap.

\begin{figure}[t]
\hspace*{-4.2mm}
\vspace*{2.5mm}
\psfragfig[scale=0.6]{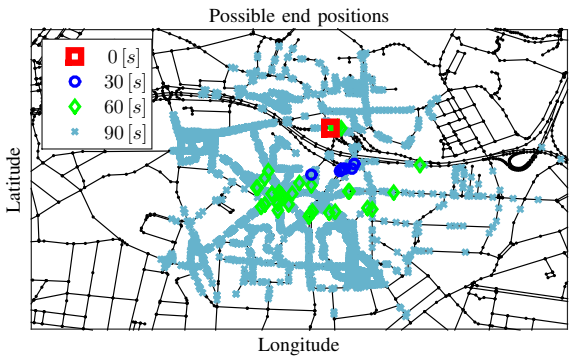}{
\psfrag{title}[][][0.75]{\raisebox{-3.5mm}{Possible end positions}}
\psfrag{ylabel}[][][0.75]{\raisebox{1.5mm}{Latitude}}
\psfrag{xlabel}[][][0.75]{\raisebox{-4mm}{Longitude}}
\psfrag{legen1}[][][0.75]{\raisebox{1.1mm}{$0\,[s]$}}
\psfrag{legen2}[][][0.75]{\raisebox{1.1mm}{$30\,[s]$}}
\psfrag{legen3}[][][0.75]{\raisebox{1.1mm}{$60\,[s]$}}
\psfrag{legen4}[][][0.75]{\raisebox{1.1mm}{$90\,[s]$}}
}
\vspace*{-7mm}
\caption{Possible positions $0\,[s]$, $30\,[s]$, $60\,[s]$, and $90\,[s]$ after initializing DR (using only measurements of speed and map information) from a known starting point. The example is taken from a trip in Porto, Portugal, and the figure shows an area of about $3.5\,[\hspace*{0.2mm}(km)^2\hspace*{0.2mm}]$.}
\label{ParticleIllustration}
\end{figure}

The DR problem arising when only speed measurements are available is fundamentally different from the DR problem when also sensor-based yaw information is available. In the latter case, the position estimate can be propagated in two dimensions using only sensor measurements. Spatial information from the road map is then conveniently incorporated into the estimation as soft constraints. Usually, this information will be sufficient to avoid filter divergence, and additional information, such as speed limits, will be of secondary importance \cite{Hall2001}, \cite{Svenzen2002}. When only measurements of speed are available, the position estimate cannot be propagated solely based on sensor measurements. Instead, the vehicle's yaw angle has to be estimated from the road map. In this case, sensor measurements and spatial information from the road map will never be sufficient as the posterior distribution will continue to "spread out" at intersections (see Fig. \ref{ParticleIllustration}). Additional models which relate the expected vehicle speed to the vehicle's position on the map are required, and speed limits will become of utmost importance.

\section{Model and estimation framework}
\label{Model_Estimation_Section}

We present a filtering-based approach to the problem of estimating a vehicle's position using map information and measurements of speed. This section describes both the employed system model and the particle-based filter implementation.

\subsection{Overview of System Model}

The map data from OpenStreetMap (OSM) can be described as a mathematical graph where each vertex (node) defines a latitude-longitude pair. The connectivity of the graph is defined by a set of ways, i.e., ordered lists of nodes indicating that there is an edge (link) between the first and the second node, between the second and the third node, etc. Each way can be characterized by attributes specifying road type, speed limits, restrictions on travel direction, etc. The problem at hand is to use information from a digital map such as OSM, together with measurements of speed, to estimate a vehicle's position. The considered estimation framework is based on the model
\begin{subequations}
\label{system_model}
\begin{align}
\mathbf{x}_{k+1}^{} &\sim p(\mathbf{x}_{k+1}^{}|\mathbf{x}_k^{},s_k^{}),\label{process_model}\\
\mathbf{y}(\mathbb{S}) &\sim p(\mathbf{y}(\mathbb{S})|\mathbb{X}).\label{meas_model}
\end{align}
\end{subequations}
Here, $\mathbf{x}$ is a state vector comprising the vehicle's two-dimensional position $\mathbf{r}$, and the vehicle's direction of travel. Assuming that the vehicle always travels along links in the graph, the state vector can be given as an ordered list of two connected nodes and a constant indicating the ratio of the vehicle's distance to the first node and the distance between the two nodes. The vehicle's speed is denoted by $s$ (no notational distinction is made between the true and measured speed). As can be seen, the speed measurements act both as input in the state equation \eqref{process_model} and as measurements in the measurement equation \eqref{meas_model}. Moreover, $p(\cdot|\cdot)$ is used to denote a conditional probability density function (pdf). All pdfs are implicitly conditioned on the available map information. Throughout the paper, we will follow the convention of using a subindex $k$ to denote quantities at sampling instance $k$. The complete set of state vectors and speed measurements are then given by
$\mathbb{X}\overset{_\Delta}{=}\{...,\mathbf{x}_{k-1}^{},\,\mathbf{x}_k^{},\,\mathbf{x}_{k+1}^{},...\}$ and
$\mathbb{S}\overset{_\Delta}{=}\{...,s_{k-1}^{},\,s_k^{},\,s_{k+1}^{},...\}$, respectively. The employed measurements are collected in a generic function $\mathbf{y}$, and will be discussed more thoroughly in Section \ref{meas_model_subsection}. For clarity, we will in the text differentiate between $s$ and $\mathbf{y}$ by referring to speed measurements and measurements, respectively.

\begin{figure}[t]
\def\svgwidth{0.8\columnwidth}
\hspace*{0mm}
\scalebox{1}{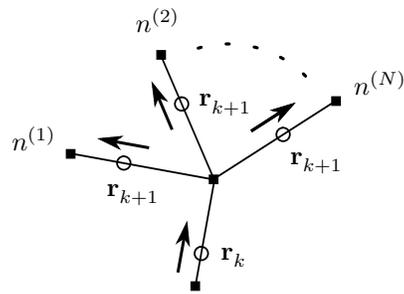}
\vspace*{-1mm}
\caption{Illustration of the process model when the vehicle passes a road junction. If the probability of the initial position $\mathbf{r}_k$ is $p$, the probability of each of the updated positions $\mathbf{r}_{k+1}^{}$ are $p/N$ where $N$ denotes the number of links along which $\mathbf{r}_{k+1}$ can be expected. The distance between $\mathbf{r}_k^{}$ and (any) $\mathbf{r}_{k+1}^{}$, measured along the links, is equal to $\Delta t_k^{}\hspace*{0.1mm}s_k^{}$.}
\label{nodes_illustration}
\vspace*{-0mm}
\end{figure}

\subsection{Process Model}
\label{process_model_subsection}

The process model \eqref{process_model} describes the time development of the vehicle's position, using speed as input. The update is performed by distributing the travelled distance $\Delta t_k^{}\hspace*{0.2mm}s_k^{}$ along the upcoming links. We have here used $\Delta t_k^{}$ to denote the sampling interval between sampling instances $k$ and $k+1$. When the vehicle passes a road junction, it is modeled to continue along any of the connected links (excluding the link it has just traversed) with equal probability. This is illustrated in Fig. \ref{nodes_illustration}. As may be realized, less probable events such as unexpected u-turns have been disregarded. If needed, the model can easily be modified to account for errors in the estimated travelled distance. Discrepancies between the estimated traveled distance $\Delta t_k^{}\hspace*{0.25mm}s_k^{}$ and the true travelled distance can arise due to map errors and off-road driving \cite{Kronander2004}. In addition,
GNSS speed measurements are subject to multipath propagation \cite{Sadrieh2012}, whereas
OBD measurements are affected by wheel slips, quantization errors in the measured speed \cite{Gustafsson2010b}, and scale factor errors dependent on tire pressure \cite{Hall2001}. Previous attempts to include the scale factor in the state vector did not result in any improvement in navigation performance, but rather, increased the risk of filter divergence \cite{Svenzen2002,Kronander2004}. We further note that the speed measured by the OBD system is the absolute value of the vehicle's three-dimensional velocity, whereas the studied model only considers two spatial dimensions (most nodes in OSM do not include any altitude information). As a result, the travelled distance in the two-dimensional plane of the map will tend to be overestimated when travelling along steep inclines \cite{Svenzen2002}. For example, if the vehicle is traveling with a speed of $25\,[\hspace*{0.2mm}m/s\hspace*{0.2mm}]$ along a road where the angle of inclination is $5\,[\hspace*{0.2mm}^{\circ}\hspace*{0.2mm}]$, the speed in the two-dimensional plane of the map will be $25\cdot\cos(5\,[\hspace*{0.2mm}^{\circ}\hspace*{0.2mm}])\,[\hspace*{0.2mm}m/s\hspace*{0.2mm}]\approx 24.62\,[\hspace*{0.2mm}m/s\hspace*{0.2mm}]$. This error contribution can be compensated for by using a commercial digital map with altitude information and extending the dimension of the state vector accordingly.

Obviously, the estimated travelled distance $\Delta t_k^{}\hspace*{0.1mm}s_k^{}$ is subject to several modeling and measurement errors. Despite this, the resulting accumulated position error will in many cases be negligible in comparison to the position error resulting from the uncertainty of the vehicle path, i.e., the list of nodes or links that the vehicle has traversed.

\subsection{Measurement Model}
\label{meas_model_subsection}

When propagating the state distribution as described in the preceding subsection, the number of feasible vehicle paths grows exponentially as a function of the traveled distance \cite{Gao2014}.
%In addition, the uncertainty of the travelled distance will increase with each sampling instance.
To bound the number of paths, measurement updates must be employed. We will give two examples of possible measurements. First, we the study the lateral forces exerted on the vehicle. When cornering, the lateral g-forces on the vehicle at sampling instance $k$ can be approximated by \cite{Wahlstrom2015}
\begin{equation}
\label{meas_two}
y^{(1)}(s_k,s_{k+1})\overset{_\Delta}{=}\sqrt{a_k^2+s_k^2\omega_k^2}
\end{equation}
where $a$ denotes the vehicle's longitudinal acceleration and $\omega$ denotes the vehicle's yaw rate. The acceleration is approximated by the difference quotient %$a_k\overset{_\Delta}{=}(s_{k+1}-s_k)/(t_{k+1}-t_k)$.
\begin{equation}
a_k\overset{_\Delta}{=}\frac{s_{k+1}-s_k}{\Delta t_k}.
\end{equation}
The yaw rate $\omega$ must be estimated from map data. We choose to estimate the yaw rate in a separate Kalman filter based on the state-space model
\begin{subequations}
\begin{align}
\theta_{k+1}&=\theta_k+\Delta t_k\hspace*{0.2mm}\omega_k, \\
\omega_{k+1}&=\omega_k+\eta_k^{\omega}, \\
\tilde{\theta}_k&=\theta_k+\epsilon_k^{\theta}.
\end{align}
\end{subequations}
Here, the noises $\eta_k^{\omega}$ and $\epsilon_k^{\theta}$ are assumed to be white with variances $\Delta t_k\sigma_{\omega}^2$ and $\sigma_{\theta}^2$, respectively. Further, $\theta$ is the vehicle's yaw angle, and $\tilde{\theta}$ is the direction of the current link. Due to the periodicity of the yaw angle, the measurement residuals must be chosen within the interval $[-180,180]\,[\hspace*{0.2mm}^{\circ}\hspace*{0.2mm}]$. An alternative to using a Kalman filter is to estimate the yaw rate directly from the difference quotients $(\tilde{\theta}_{k+1}-\tilde{\theta}_k)/\Delta t_k$.
%of travel directions (as given by the map at the current and previous position) and the corresponding time instances.
However, due to the limited resolution of the nodes in OSM (many sharp corners will have links meeting at a $90\,[\hspace*{0.2mm}^{\circ}\hspace*{0.2mm}]$ angle) these difference quotients will be fairly unstable.

Typically, passenger cars are not designed for g-forces higher than $0.8\,g$, and will seldom exceed $0.6\,g$ during normal driving. (A sufficiently high lateral force will cause the vehicle to rollover \cite{Wahlstrom2015}.) Here, $g$ denotes the gravitational acceleration at the surface of the earth. Motivated by this, the distribution of $y^{(1)}$ is heuristically modeled as
\begin{align}
\begin{split}
& p(y^{(1)}(s_k,s_{k+1})|\mathbf{x}_{1},...,\mathbf{x}_{k})\\[1.5ex]\propto&
\begin{cases}
1, & y^{(1)}<g^{(1)} \vspace*{0.7mm} \\
\frac{\displaystyle y^{(1)}-g^{(2)}}{\displaystyle g^{(1)}-g^{(2)}}, & g^{(1)}\le y^{(1)} < g^{(2)} \vspace*{0.7mm} \\
0, & g^{(2)}\le y^{(2)}
\end{cases}.
\end{split}
\end{align}
The design parameters $g^{(1)}$ and $g^{(2)}$ are chosen so that the lateral forces fall below $g^{(1)}$ during the larger part of the driving while allowing for outliers giving estimated lateral forces up to $g^{(2)}$.

To derive the second type of measurements, we note that drivers tend to maintain a speed close to the speed limit for the larger part of the route. Hence, we propose the use of the measurements
%\begin{subequations}
%\begin{align}
%g_1(y_k,y_{k+1})&\overset{_\Delta}{=}\frac{y_{k+1}-y_k}{\Delta T_k}+c\hspace*{0.2mm}y_k\\
%y_0&\overset{_\Delta}{=}\sum_{i=0}^{K}w^{(i)}s(\mathbf{x}_{k+i})
%\end{align}
%\end{subequations}
\begin{equation}
\label{meas_one}
y^{(2)}(s_k,s_{k+1})\overset{_\Delta}{=}a_k+c\hspace*{0.2mm}s_k
\end{equation}
distributed according to
\begin{equation}
\label{distribution_g1}
p(y^{(2)}(s_k,s_{k+1})|\mathbf{x}_1,...,\mathbf{x}_{k+K}) = p_{\cal C}(y^{(2)},c\hspace*{0.3mm}\bar{s}_{k+K},\sigma)
\end{equation}
where $K\in\mathbb{N}$ and $p_{\cal C}$ denotes the pdf of the Cauchy distribution with location and scale parameters given by the second and third arguments, respectively. The probabilistic assumption in \eqref{distribution_g1} is derived from the model
\begin{equation}
\label{acc_model}
a_k=c\hspace*{0.3mm}(\bar{s}_{k+K}-s_k)+\epsilon_k^s
\end{equation}
which says that the vehicle's acceleration will be approximately proportional to the difference between the current speed and the target speed $\bar{s}_{k+K}$. As should be obvious, we have assumed that $\epsilon_k^s$ is Cauchy distributed with location $0$ and scale $\sigma$. The design parameters $c$ and $\sigma$ describe the characteristics of each individual driver and car. Modeling the error $\epsilon$ as distributed according to the the heavy-tailed Cauchy distribution will make the estimation robust to measurement errors caused by unexpected driving behavior.

The target speed is, just as the yaw rate, estimated in a separate Kalman filter. In this case, the state-space model is
\begin{subequations}
\begin{align}
\bar{s}_{k+1}&=\bar{s}_k+\eta_k^s. \\
\tilde{s}_k&=\bar{s}_k+\epsilon_k^s.
\end{align}
\end{subequations}
Here, the noises $\eta_k^s$ and $\epsilon_k^s$ are white with variances $\Delta t_k\sigma_{s1}^2$ and $\sigma_{s2}^2$, respectively, while $\tilde{s}$ is the speed limit of the current link. By letting the target speed be dependent on the speed limits on the road ahead, we are able to capture the forward-looking behavior of typical driving. At each iteration, the estimated target speed is projected to the closest speed in the space
\begin{equation}
\{s_k:\sqrt{a_k^2+s_k^2\omega_k^2}<g^{(1)}\}.
\end{equation}
In other words, the estimated target speed is adjusted so as to never imply an excessively large lateral force. This means that particles which pass sharp corners while the vehicle slows down often will have their weights increased, and hence, we can detect the vehicle turning even though all nearby links have the same speed limit.

As an alternative to using the model \eqref{acc_model}, it is possible to derive a measurement function based on the assumption that the current speed is close to the target speed. This model is obtained as a special case of \eqref{acc_model} in the limit of $\bar{s}_{k+K}=s_k$. However, this model tends to be too restrictive on discrepancies between the vehicle speed and the target speed, and often causes the measurement errors to have a significant temporal correlation (e.g., if the vehicle's speed exceeds the target speed at one sampling instance it is very likely to do so also at the next sampling instance). By using \eqref{acc_model}, we allow for temporally correlated discrepancies between the vehicle speed and the target speed, but still expect the speed to be at least approaching the target speed.

\algrenewcommand\algorithmicindent{1.2em}%
\begin{algorithm}[t]
\caption{\vspace*{-0.55mm}: Particle-filter algorithm. \label{alg1}}
\begin{algorithmic}[1]
\vspace*{0.4mm}
\For{$k=0,1,\dots$} \vspace*{0.4mm}
\State \emph{Time update}: Update each particle state $\mathbf{x}_k$ and its associated particle weight $w_k$ by distributing the travelled distance $\Delta t_{k}^{}\hspace*{0.2mm}s_{k}^{}$ along the upcoming links.
\State \emph{Parameter update}: Perform a time update in the Kalman filters estimating the yaw angle, the yaw rate, and the target speed for each particle. Use the travel direction and the speed limit at each particle to perform the associated measurement updates.
\State \emph{Measurement update}: Update the particle weights according to equation \eqref{part_weight_upd_eq}. The estimated yaw rate is used in the updates based on both $y^{(1)}$ and $y^{(2)}$, whereas the estimated target speed only is needed in the update based on $y^{(2)}$. Set the weights of all particles which violate constraints on travel directions to zero.
\State \emph{Particle weight redistribution}: Redistribute the prior weight of particles which have had their weight set to zero in the measurement update to particle "siblings" (if such siblings exists).
\State \emph{Particle elimination}: Eliminate all particles with zero weight. If the total number of particles exceeds a given threshold, eliminate all particles with insufficient weight.
\State \emph{Particle resampling}: If $k=n\cdot N_0$ for some $n\in\mathbb{N}$ and some integer parameter $N_0$, and if the number of particles is sufficiently low, resample particles using systematic resampling. Displace each sampled particle along its current link by sampling from a uniformly distributed distance. Re-initialize the particles so that all particle weights are equal.
\State \emph{Particle merging}: Merge particles that are sufficiently similar.
\EndFor
\end{algorithmic}
\end{algorithm}

\begin{table}[t]
\vspace*{-3mm}
\begin{tabular}{l}
\hspace*{-2mm}{\footnotesize{The algorithm is detailed in Sections \ref{process_model_subsection}, \ref{meas_model_subsection}}, and \ref{particle_filter_subsection}.}
\end{tabular}
\vspace*{-4mm}
\end{table}

Both types of measurements $y^{(1)}$ and $y^{(2)}$ are based on ideas similar to those in \cite{Gao2014}. Additional measurements can be derived from the fact that vehicle stops tend to coincide with traffic lights or stop signs in the map. This information was not utilized here as the number of unpredictable stops (e.g., stops due to congestion) was very large, and information on traffic lights and stop signs was lacking over the larger part of the map area.

Since all speed measurements are used both in the process model \eqref{process_model} and in the measurements \eqref{meas_two} and \eqref{meas_one} one might argue that the system model includes several error correlations that must be considered in the estimation. However, the larger part of the variance in $y^{(1)}(s_k,s_{k+1})|\mathbf{x}_{1},...,\mathbf{x}_{k}$ and $y^{(2)}(s_k,s_{k+1})|\mathbf{x}_{1},...,\mathbf{x}_{k+K}$ stem from variations in driving behavior, rather than from errors in the measured speed (for example, the absolute OBD measurement error is typically smaller than $1\,[km/h]$, while the speed at a given position often can differ by $20\,[km/h]$ depending on the traffic conditions). As a result, the correlations caused by errors in the measured speed are generally negligible.

\subsection{Particle Filter Implementation}
\label{particle_filter_subsection}

Due to the multimodal characteristics of the conditional pdfs in the system model \eqref{system_model}, assuming unimodal posterior distributions will typically lead to irrecoverable localization errors in a very short period of time. To avoid this, we propose a solution based on the particle filter framework \cite{Gordon1993}. Each particle must, in addition to the current state vector $\mathbf{x}_k$, also store its estimates of the yaw angle, the yaw rate, and the target speed. However, as the state-space models for these parameters are time-invariant (up to variations in the sampling interval), the covariance matrices only need to be stored once. The particle propagation is performed according to the model described in Section \ref{process_model_subsection}. Instead of sampling a single continuing path when a particle passes a road junction (i.e., ignoring all but one of the possible links), the particle is split into several new particles with equal weights (the sum of which is equal to the weight of the original particle). In this way, we increase the particle diversity and decrease the risk of filter divergence due to unlucky sampling. Obviously, this also means that the number of particles in many cases will increase during the propagation step.

Aside from the resulting increase in computational cost, the choice of "splitting" particles as they pass a road junction has an additional downside. This becomes clear when the true path of the vehicle passes a large number of road junctions. In this case, a particle following this path will have its weight diminished with each passing road junction. If the weight of the particle is not to fall short of resampling thresholds, it must then increase substantially during measurement updates. In practice, this means that particles often will be eliminated mainly because they passed many road junctions. To mitigate this effect, we once again utilize the fact that the sum of the weights of particles emerging from a particle passing a road junction should be equal to the weight of the original particle.
%In other words, the probability that a vehicle turned left or right at a three-way junction is equal to the probability that the vehicle passed the junction in the first place.
Hence, if a particle which has emerged from a "particle split" at a road junction is immediately eliminated due to an estimated lateral force exceeding $g^{(2)}$, its weight, prior to the elimination, is distributed to its "siblings", i.e., other particles which originated from the same road junction and particle. The same rules are applied to particles which violate constraints on travel directions, i.e., which travel in the wrong direction on a one-way link. Implementing this modification will tend to reduce the risk of filter divergence when the vehicle travels along a main road with a large number of intersections.

As motivated in the last paragraph in Section \ref{process_model_subsection}, the particle propagation is performed without any simulated process noise in the travelled distance. Similar modeling choices have previously been discussed in \cite{Nilsson2013} and references therein.

%It should, however, be noted that accumulated errors in the travelled distance will be observable since particles emerging from the resampling step are spread out by random displacements (see below).

In the measurement update, each particle weight $w_k$ is updated according to
\begin{equation}
\label{part_weight_upd_eq}
w_{k+1}\propto p(\mathbf{y}_{k+1}(\mathbb{S})|\mathbb{X})\hspace*{0.2mm}w_k
\end{equation}
where the total set of measurements are
\begin{equation}
\mathbf{y}_{k+1}(\mathbb{S})\overset{_\Delta}{=}\big[\begin{matrix}
y^{(1)}(s_k,s_{k+1}) & y^{(2)}(s_{k-K},s_{k+1-K})
\end{matrix}\big]^{\intercal}.
\end{equation}
Note that the measurements $y^{(2)}$ use earlier speed measurements than $y^{(1)}$ as the conditional distribution of $y^{(2)}(s_{k-K},s_{k+1-K})$ is modeled as dependent on the target speed at sampling instance $k$.

The filter is implemented with systematic resampling \cite{Hol2004}. New particles are resampled at equidistant points in time given that the number of particles is sufficiently low. Each new particle is displaced with a uniformly distributed distance along its trajectory. This will make it possible to detect any eventual accumulated errors in the total travelled distance.
Particles with insufficient weight are eliminated whenever the total number of particles exceed a given threshold. However, particles which are given a zero weight are eliminated immediately. In addition, whenever two particles are sufficiently similar (i.e., are on the same link, have the same parameter estimates, and have sufficiently similar position estimates) they are merged into a single particle which will have its weight equal to the sum of the original weights and its position equal to the weighted average of the original positions. This will ease the computational burden somewhat as well as mitigate problems with particle clustering \cite{Hall2001}. The particle filter is summarized in Algorithm \ref{alg1}.

\begin{table}[t]
\caption{filter parameters. \label{TabParameters}}
\centering
\begin{tabular}{ll|r}
\hline
\hline
& & \\ [-2.3ex]
& Parameter & Value \\
& & \\ [-2.7ex]
\hline
& & \\ [-2.3ex]
\emph{Process noise} & $\sigma_\omega$  & $5\,[\hspace*{0.2mm}^{\circ}/s^2/\sqrt{H\!z}\hspace*{0.2mm}]$ \\ [-0.2ex]
& $\sigma_{s1}$ & $0.5\,[\hspace*{0.2mm}m/s^2/\sqrt{H\!z}\hspace*{0.2mm}]$ \\ [-2.6ex]
& & \\
\hline
& & \\ [-2.5ex]
\emph{Measurement noise} & $\sigma_\theta$  & $15\,[\hspace*{0.2mm}^{\circ}\hspace*{0.2mm}]$ \\ [-0.2ex]
& $\sigma_{s2}$  & $10\,[\hspace*{0.2mm}m/s\hspace*{0.2mm}]$ \\ [-2.6ex]
& & \\
\hline
& & \\ [-2.4ex]
$y^{(1)}$ \emph{update} & $g^{(1)}$  & $0.55\,g$ \\ [-0.2ex]
& $g^{(2)}$  & $0.65\,g$ \\ [-2.6ex]
& & \\
\hline
& & \\ [-2.4ex]
$y^{(2)}$ \emph{update} & $K$  & $12$ \\ [-0.2ex]
& $c$  & $0.05\,[\hspace*{0.2mm}1/s\hspace*{0.2mm}]$ \\ [-0.2ex]
& $\sigma$  & $1.5\,[\hspace*{0.2mm}m/s^2\hspace*{0.2mm}]$ \\ [-2.6ex]
& & \\
\hline
& & \\ [-2.4ex]
\multicolumn{2}{l|}{Particle threshold (resampling trigger)$^{\dagger}$} & $100$ \\ [-0.2ex]
\multicolumn{2}{l|}{Weight threshold (elimination)} & $1/200$ \\ [-0.2ex]
\multicolumn{2}{l|}{Distance threshold (particle merge)} & $20\,[m]$ \\ [-0.2ex]
\multicolumn{2}{l|}{Resampling displacement interval} & $(-100,100)\,[m]$ \\ [-2.6ex]
& & \\
\hline \hline
\multicolumn{3}{l}{} \\ [-2.5ex]
\multicolumn{3}{l}{\footnotesize{$^{\dag}$ Checked every hundredth second.}} \\ [-0.2ex]
\end{tabular}
\vspace*{-0mm}
\end{table}

\section{Field study}
\label{Field_Study_Section}

GNSS and OBD data were collected from five drivers performing a total of eighteen different trips in both rural and urban areas in and around Porto, Portugal. No specific trip was conducted for the purpose of this study, and hence, the data represents a random sample of typical routes in the area (however, for practical reasons, the trips were somewhat skewed towards more lengthy trips). All data was collected using the SenseMyCity app, which was developed as part of the Future Cities project \cite{Rodrigues2014}. The median and mean trip lengths were $28.3\,[km]$ and $49.5\,[km]$, respectively. The update date rate of the GNSS data was $1\,[H\!z]$, while the update rate of the OBD data varied between $1\,[H\!z]$ and $3\,[H\!z]$ depending on the car (the majority of the trips had an update rate close to $1\,[H\!z]$). Using a dedicated setup collecting OBD speeds with a higher update rate may improve the estimation performance.
%The proposed estimation framework was evaluated using speed measurements from both smartphone-embedded GNSS receivers and the OBD system.
The GNSS position measurements were used as ground truth. Assuming knowledge of the vehicle's initial position (this will often be the driver's home address, the driver´s work address, etc.), the filter was initialized by sampling particles on all links in a circle of radius $50\,[m]$ centered at the true initial position. Due to the stochastic nature of the filter, all displayed performance measures have been averaged over 10 runs.
The parameters used in the field study are shown in Table \ref{TabParameters}.

Occasionally, data was lost from either of the two sensors. In these cases, the missing data was replaced by data from the other sensor (GNSS replacing OBD or OBD replacing GNSS). The replacement data was scaled to take the OBD scale factor into account. The scale factor can be estimated using data from sampling instances where measurements from both sensors are available. While some GNSS outages always can be expected, we stress that the larger part of the data collection disruptions occurred in the OBD data and was related to the chosen communications setup. In other words, data losses of this kind should not be expected in e.g., a commercial OBD-based insurance telematics program. In two (five) of the trips, data losses caused the measurements of the trip to start (end) at nonzero speeds. If the data starts or ends close to the motorway, this will usually simplify the estimation since filter divergence is more prone to occur along roads with lower speed limits.

\subsection{Results}

\begin{figure}[t]
\hspace*{-5mm}
\vspace*{2.5mm}
{
\psfragfig[scale=0.6]{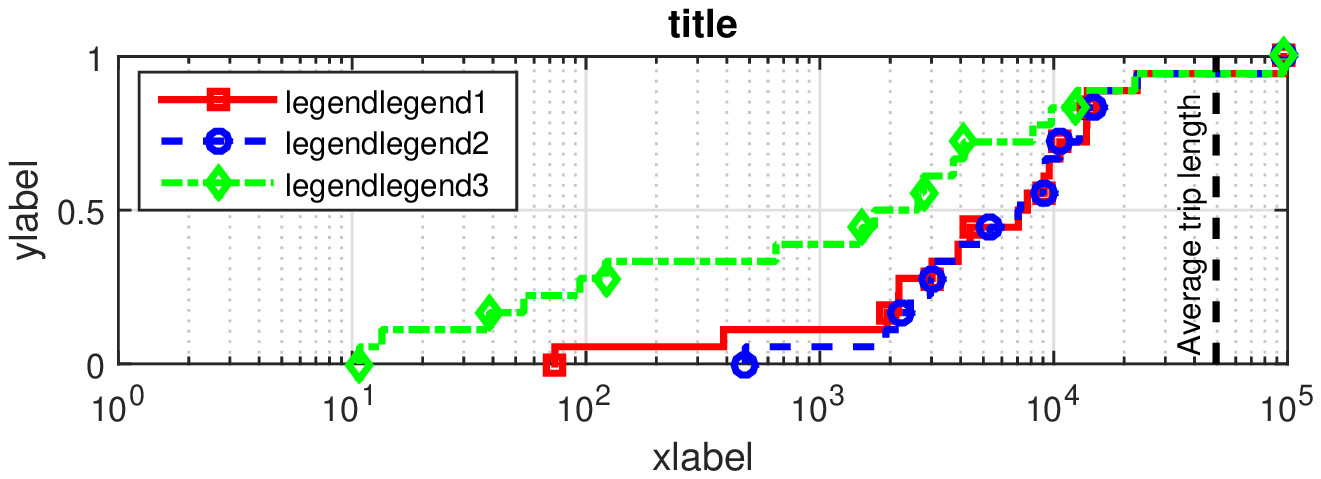}{
\psfrag{xlabel}[][][0.75]{\raisebox{-3.5mm}{$\Vert\hat{\mathbf{r}}-\mathbf{r}\Vert\,[m]$}}
\psfrag{ylabel}[][t][0.75]{edf}
\psfrag{title}[][][0.75]{\raisebox{-3.5mm}{(a) Horizontal position error}}
\psfrag{legendlegend1}[][][0.75]{\raisebox{0.9mm}{MAP}}
\psfrag{legendlegend2}[][][0.75]{\raisebox{0.9mm}{MMSE}}
\psfrag{legendlegend3}[][][0.75]{\raisebox{0.9mm}{BP}}
}
}
\hspace*{-5mm}
{
\psfragfig[scale=0.6]{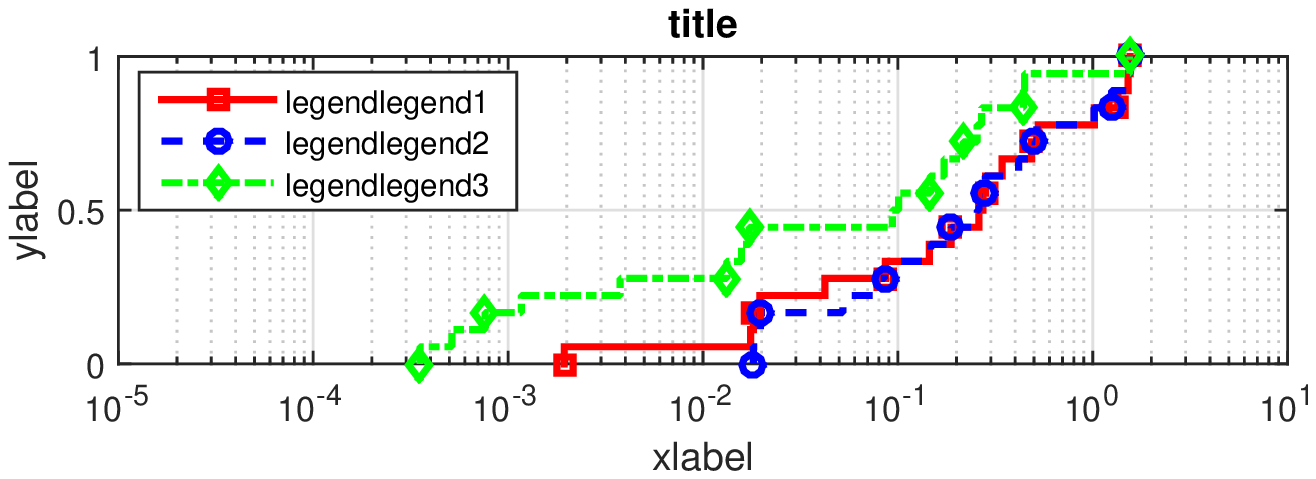}
{
\psfrag{xlabel}[][][0.75]{\raisebox{-3.5mm}{$\Vert\hat{\mathbf{r}}-\mathbf{r}\Vert/L$}}
\psfrag{ylabel}[][t][0.75]{edf}
\psfrag{title}[][][0.75]{\raisebox{-3.5mm}{(b) Relative horizontal position error}}
\psfrag{legendlegend1}[][][0.75]{\raisebox{0.9mm}{MAP}}
\psfrag{legendlegend2}[][][0.75]{\raisebox{0.9mm}{MMSE}}
\psfrag{legendlegend3}[][][0.75]{\raisebox{0.9mm}{BP}}
}
}
\vspace*{-7mm}
\caption{The empirical distribution function of the horizontal and relative horizontal position errors of the end position. The filter used GNSS measurements of speed.}
\label{ErrFigGNSS}
\vspace*{-1mm}
\end{figure}

\begin{figure}[t]
\hspace*{-5mm}
\vspace*{2.5mm}
{
\psfragfig[scale=0.6]{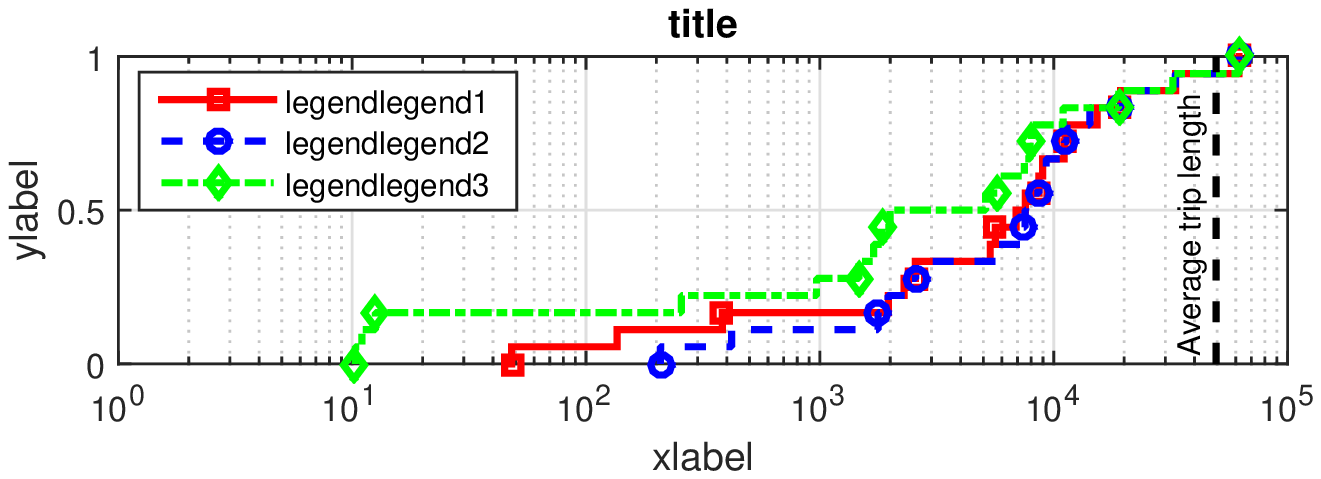}{
\psfrag{xlabel}[][][0.75]{\raisebox{-3.5mm}{$\Vert\hat{\mathbf{r}}-\mathbf{r}\Vert\,[m]$}}
\psfrag{ylabel}[][t][0.75]{edf}
\psfrag{title}[][][0.75]{\raisebox{-3.5mm}{(a) Horizontal position error}}
\psfrag{legendlegend1}[][][0.75]{\raisebox{0.9mm}{MAP}}
\psfrag{legendlegend2}[][][0.75]{\raisebox{0.9mm}{MMSE}}
\psfrag{legendlegend3}[][][0.75]{\raisebox{0.9mm}{BP}}
}
}
\hspace*{-5mm}
{
\psfragfig[scale=0.6]{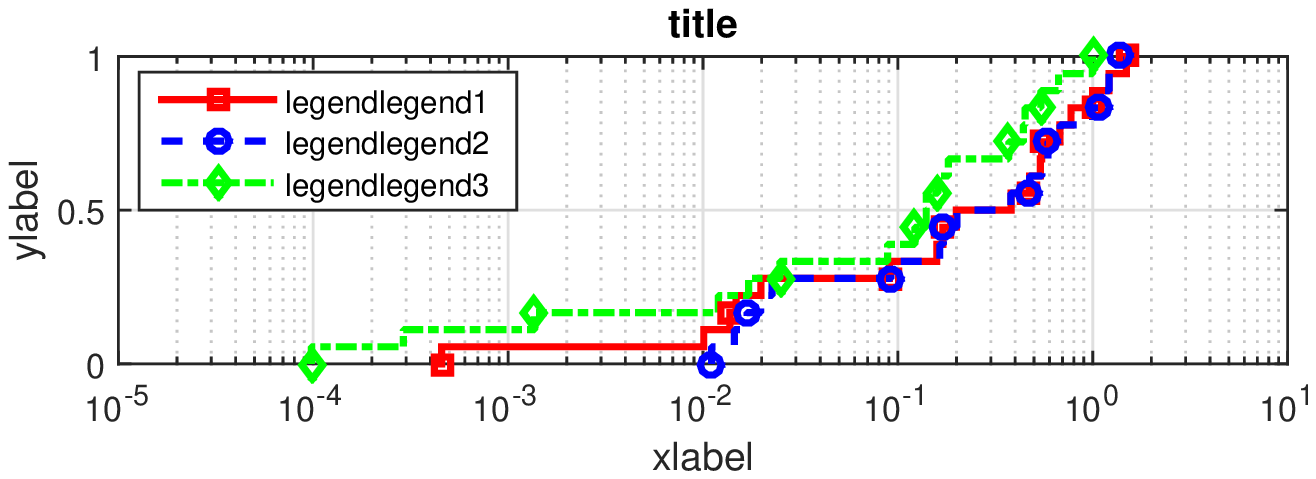}
{
\psfrag{xlabel}[][][0.75]{\raisebox{-3.5mm}{$\Vert\hat{\mathbf{r}}-\mathbf{r}\Vert/L$}}
\psfrag{ylabel}[][t][0.75]{edf}
\psfrag{title}[][][0.75]{\raisebox{-3.5mm}{(b) Relative horizontal position error}}
\psfrag{legendlegend1}[][][0.75]{\raisebox{0.9mm}{MAP}}
\psfrag{legendlegend2}[][][0.75]{\raisebox{0.9mm}{MMSE}}
\psfrag{legendlegend3}[][][0.75]{\raisebox{0.9mm}{BP}}
}
}
\vspace*{-7mm}
\caption{The empirical distribution function of the horizontal and relative horizontal position errors of the end position. The filter used OBD measurements of speed.}
\label{ErrFigOBD}
\end{figure}

The performance of the filter when using GNSS and OBD measurements of speed are shown in Fig. \ref{ErrFigGNSS} and Fig. \ref{ErrFigOBD}, respectively. The figures display the empirical distribution functions (edfs) of the horizontal $\Vert\hat{\mathbf{r}}-\mathbf{r}\Vert$ and relative horizontal position errors $\Vert\hat{\mathbf{r}}-\mathbf{r}\Vert/L$ of the estimated end position. Here, we have used $L$ to denote the driving length of the trip. If all particles were eliminated while running the filter, the horizontal position error was considered to be equal to the driving length of the trip. The performance of two estimators are shown. These are the maximum a posteriori (MAP) estimator, i.e., the position of the particle with the largest weight, and the minimum mean square error (MMSE) estimator, i.e., the weighted average of all particle's end points. In addition, we also display the accuracy of the particle with the position estimate that is closest to the true end position. This is called the best-particle (BP). Obviously, this is not a realizable estimator as we can only find the BP by assuming knowledge of the true end position. However, we still believe the accuracy of the BP is relevant as it gives an indication of the performance that can be expected when additional information on likely end positions are available (whereas the MAP and MMSE estimators give an indication of the performance that can be expected when processing trips independently without access to any additional personal information). Many drivers tend to start and end most of their trips at a very limited number of locations. Hence, when a vehicle is detected to have stopped for an extended period of time, the current particles can be evaluated based on personal information, previously extracted driving habits, and map information (for example, a driver will seldom stop in the middle of the motorway). Under ideal circumstances, only one particle will be considered to be a realistic end point.

\begin{figure}[t]
\hspace*{-5mm}
\vspace*{2.5mm}
{
\psfragfig[scale=0.6]{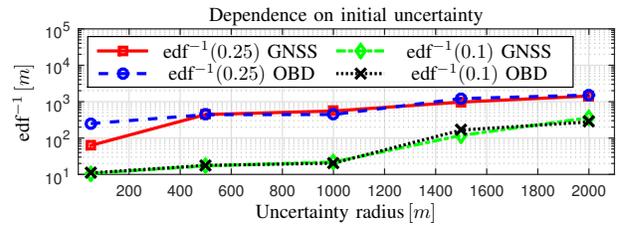}{
\psfrag{xlabel}[][][0.75]{\raisebox{-3mm}{Uncertainty radius$\,[m]$}}
\psfrag{ylabel}[][t][0.75]{edf$^{-1}\,[m]$}
\psfrag{title}[][][0.75]{\raisebox{-3.5mm}{Dependence on initial uncertainty}}
\psfrag{legend1}[][][0.75]{\hspace*{23mm}\raisebox{1.4mm}{edf$^{-1}(0.25)$ GNSS}}
\psfrag{legend2}[][][0.75]{\hspace*{23mm}\raisebox{1.4mm}{edf$^{-1}(0.25)$ OBD}}
\psfrag{legend3}[][][0.75]{\hspace*{23mm}\raisebox{1.4mm}{edf$^{-1}(0.1)$ GNSS}}
\psfrag{legend4}[][][0.75]{\hspace*{23mm}\raisebox{1.4mm}{edf$^{-1}(0.1)$ OBD}}
}
}
\vspace*{-8mm}
\caption{The empirical $0.1$ and $0.25$ quantiles of the horizontal position error of the end position of the BP, as dependent on the radius of the circle in which the initial particles were sampled.}
\label{initDependence}
\end{figure}

As can be seen, the BP lies less than $200\,[m]$ from the end position in about a fourth of the trips, and lies less than $2\,[km]$ from the end position in about half of the trips. If the true end position is e.g., the driver's home, this level of accuracy will often be sufficient to determine that the driver went home during the trip. Similarly, the MAP estimator is accurate to within approximately $200\,[m]$ in a tenth of the trips. Moreover, the errors of the MAP and MMSE estimators are about a fifth of the total trip length in about half of the trips. This means that even though we are not be able to accurately estimate the end position of a trip, we might still be able to estimate the overall direction of travel (e.g., whether the driver drove towards or away from the city center).

The dependence on the initial uncertainty is illustrated in Fig. \ref{initDependence} which shows the quantiles edf$^{-1}(0.1)$ and edf$^{-1}(0.25)$ of the BP estimate as dependent on the uncertainty radius, i.e., the radius of the circle in which the initial particles were sampled. All other filter parameters were kept constant. Even with an uncertainty radius of $1\,[km]$, a fourth of the trips have a horizontal position error below $600\,[m]$. This demonstrates the capability to extract positional information from speed measurements even when there is significant uncertainty in the inital position.

The estimated OBD scale factor had a sample mean and sample standard deviation (over the studied trips) of $1.012$ and $0.024$, respectively. Despite this, the choice of using GNSS or OBD measurements of speed seem to have a limited effect on the filter performance.

\subsection{Filter Divergence}

When no particle is closer than approximately $200\,[m]$ to the true position, the filter should be considered to have diverged. In other words, the filter ultimately diverged in a majority of the trips studied here. However, in many cases the the filter did not diverge until late into the trip, and hence, the error of the end position could still be small as compared to the total driving length. The filter will usually diverge in urban environments where the road density is high and most roads have the same speed limit. In these cases, the number of possible paths grows very fast and the measurement updates provide little information. While the rate of divergence may seem rather high, the results (e.g., an MMSE estimator with a median horizontal error of about a fifth of the driving length) are comparable to those obtained in \cite{Gao2014} where the median trip length was $7.5\,[km]$ (about four times shorter than in the data used here), the maximum trip length was $16.0\,[km]$ (shorter than both the median and mean trip length in the data used here), and no uncertainty in the initial position was mentioned. It can also be noted that filter divergence occasionally can occur even if measurements of yaw rate are available \cite{Hall2001,Svenzen2002,Hedlund2008}.

\section{Discussion}
\label{discussion_section}

\begin{table}[t]
\caption{capability of proposed method for navigation.  \label{FilterCapabilityTab}}
\centering
\begin{tabular}{l|c}
\hline
\hline
& \\ [-2.3ex]
Inference of end destination & Share of trips \\
& \\ [-2.7ex]
\hline
& \\ [-2.3ex]
Exact location ($\sim 200\,[m]$) & $1/4$ \\ [-0.2ex]
Approximate region ($\sim 2\,[km]$) & $1/2$ \\ [-2.6ex]
& \\
\hline \hline
\end{tabular}
\vspace*{-0mm}
\end{table}

In the preceding section, digital map information and measurements of speed were used to geographically locate a car during everyday trips. The estimates were shown to be accurate enough to enable violations of the driver's locational privacy in a substantial amount of the trips. (A summary of the indicated capability is provided in Table \ref{FilterCapabilityTab}.) As a result, this also means that the data collected in many of the current insurance telematics programs is far more sensitive than stated by the insurance companies themselves. While many of these companies will have privacy policies that regulate their data practices, the data can easily be misused if it falls into the wrong hands. Moreover, since the privacy risks will not always be obvious for someone without experience in navigation, the data might not be treated as sensitive, thereby increasing the risk of data theft. Normally, the data is saved indefinitely without consideration of any "right to be forgotten".

Insurance companies offering telematics services will often have years' worth of data attributed to individual policyholders, and can therefore make use of previously extracted personal information (e.g., commonly visited locations) to improve the position estimation over time. One way to do this could be to apply more sophisticated weight distributions in the propagation step presented in Section \ref{process_model_subsection}. Put differently, the weights of different particles emerging from the crossing of a road junction could be adjusted based on previous inferences on driving habits.

The prospect of being able to extract positional information from speed measurements does not only pose privacy risks, but also means that insurance companies can use a larger number of features to distinguish and characterize policyholders. For example, while speed measurements alone only enable studies of absolute speeding, i.e., whether the driver exceeds a given speed threshold, accurate knowledge of both speed and position makes it possible for the insurer to directly identify eventual speed limit violations. In addition, the insurer can penalize driving in areas where accidents are prone to happen.

There are several ways to implement a privacy-preserving insurance telematics program. For example, the data assembled at the central server does not need to reveal from which vehicle a specific set of data was gathered. This approach is taken in \cite{Popa2009}, which associates each collected data tuple with a time stamp and a random time-varying vehicle identifier. Any information extraction (e.g., the computation of a driver score) requiring data from a single vehicle must then be performed as a secure multi-party computation involving both the server and the vehicle-fixed client application. In this case, ideal privacy preservation means that the server data by itself cannot be used to extract more information about an individual vehicle's data than can be done from the same set of data but without any vehicle identifiers. However, if position data is collected, it may still be possible to identify data from policyholders who live in secluded areas (since they might be the only people driving in this area) \cite{Hoh2006, Krumm2007}. The corresponding location traces may then be found by using that location updates close in time and space are likely to originate from the same vehicle \cite{Gruteser2005}. Further, location traces from different drivers can be clustered by studying individual driving characteristics, so called driver fingerprinting \cite{Enev2016}. Obviously, it is possible to reduce the risk of inference attacks by simply reducing the data quality, assuming that the primary privacy-preserving information extraction can still be performed with tolerable errors. This can be done by e.g., omitting samples or deliberately perturbing measurements \cite{Hoh2007}.

Many insurance telematics providers are not interested in speed measurements or location updates per se, but rather, in the driver score or risk profile that can be extracted from them. Hence, it will often be sufficient to compute the measures of interest directly on the telematics unit, send these to a central server, and discard the original measurements \cite{Handel2014_2}. This will both reduce the required communications and the risks of privacy intrusion. On the downside, this may increase the computational demands on the telematics units.

%However, on the downside, this approach will make it harder to utilize the raw measurements for value-added services based on e.g., traffic state estimation or road condition monitoring, and it will also increase the computational demands on the telematics units.

%\cite{Wolf2014}

\section{Conclusions}
\label{conclusions_section}

We have considered the problem of estimating the position of a vehicle using only map information and measurements of speed. As opposed to previous studies utilizing the same set of sensors, the problem was here formulated within the well-known particle filter framework. Particle-based estimation makes it straightforward to 1) study the estimation accuracy using estimators suited for multimodal posteriors; 2) draw use of experience from previous filter-based approaches where sensor-based yaw information have been available; and 3) extend the presented framework with additional measurement functions. Performance evaluations of the proposed method indicated that location-based information such as end destinations often can be estimated with an accuracy in the order of $100\,[m]$. This level of accuracy will in many cases be sufficient to extract information on e.g., visited locations or areas. As a result, the collection of speed measurements from policyholders in current insurance telematics programs both poses privacy risks and makes it possible for the insurer to discriminate among drivers based on visited areas. Telematics providers who wishes to decrease the risk of privacy infringement for their policyholders can for example make use of time-varying vehicle identifiers, or perform the computations of relevant driver measures directly in the vehicle-fixed telematics units.

%undisclosed
%privacy infringement

% Can use something like this to put references on a page
% by themselves when using endfloat and the captionsoff option.
\ifCLASSOPTIONcaptionsoff
  \newpage
\fi

% trigger a \newpage just before the given reference
% number - used to balance the columns on the last page
% adjust value as needed - may need to be readjusted if
% the document is modified later
%\IEEEtriggeratref{8}
% The "triggered" command can be changed if desired:
%\IEEEtriggercmd{\enlargethispage{-5in}}

% references section

% can use a bibliography generated by BibTeX as a .bbl file
% BibTeX documentation can be easily obtained at:
% http://www.ctan.org/tex-archive/biblio/bibtex/contrib/doc/
% The IEEEtran BibTeX style support page is at:
% http://www.michaelshell.org/tex/ieeetran/bibtex/
\bibliographystyle{IEEEtran}
% argument is your BibTeX string definitions and bibliography database(s)
\bibliography{refs}

\begin{IEEEbiography}[{\includegraphics[width=1in,height=1.25in,clip,keepaspectratio]{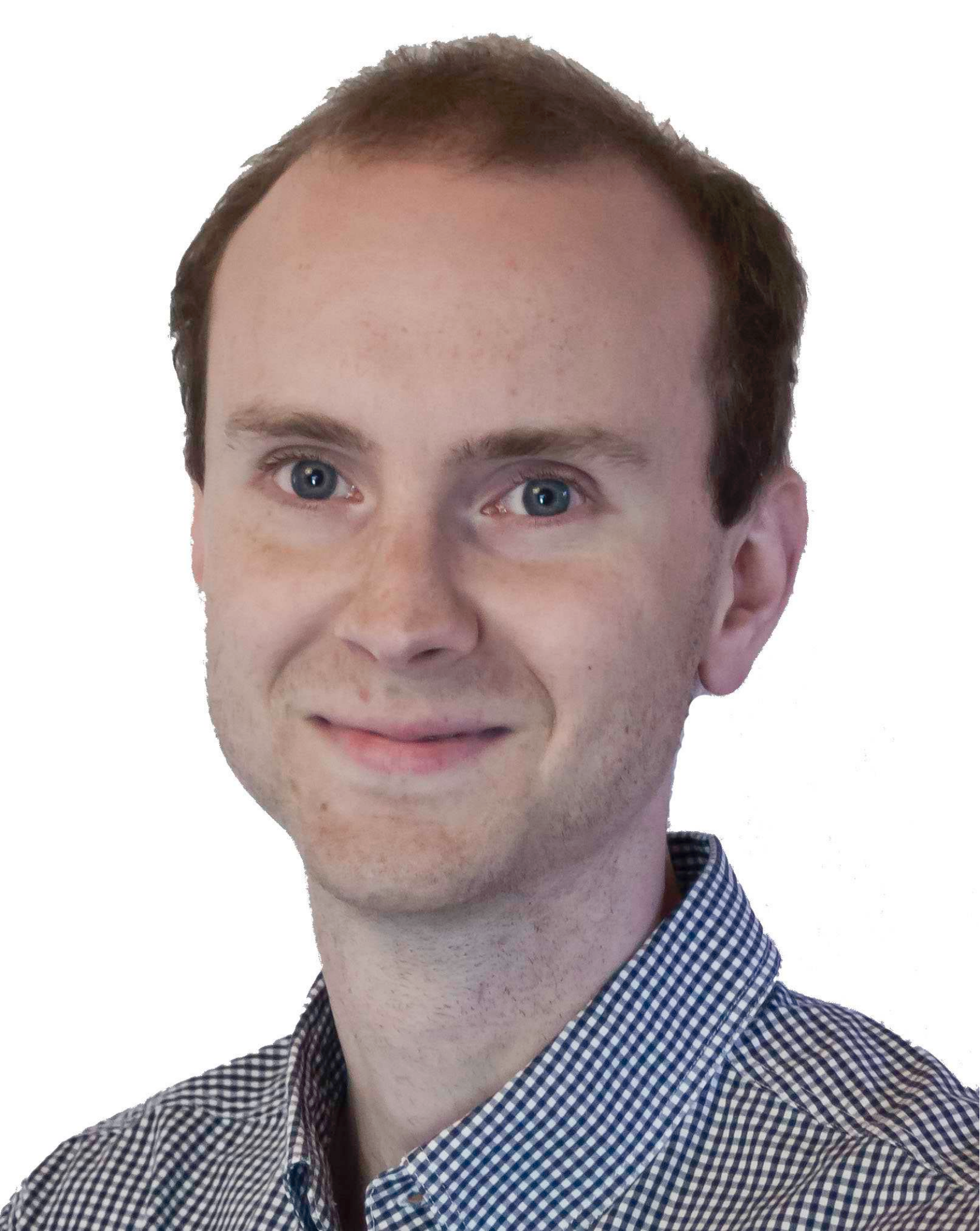}}]
{Johan Wahlström} received his MSc degree in Engineering Physics from the KTH Royal Institute of Technology, Stockholm, Sweden, in 2014. He subsequently joined the Signal Processing Department at KTH, working towards his PhD. His main research topic is insurance telematics. In 2015, he received a scholarship from the Sweden-America foundation and spent six months at Washington University, St. Louis, USA.
\end{IEEEbiography}

\begin{IEEEbiography}[{\includegraphics[width=1in,height=1.25in,clip,keepaspectratio]{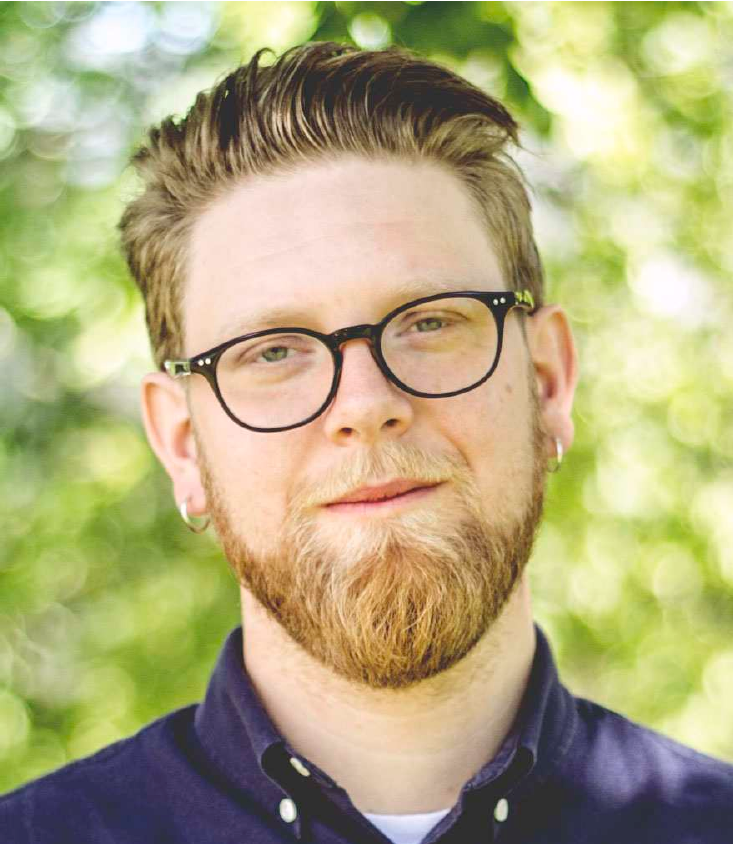}}]
{Isaac Skog}(S'09-M'10) received the BSc and MSc degrees in Electrical Engineering from the KTH Royal Institute of Technology, Stockholm, Sweden, in 2003 and 2005, respectively. In 2010, he received the Ph.D. degree in Signal Processing with a thesis on low-cost navigation systems. In 2009, he spent 5 months at the Mobile Multi-Sensor System
research team, University of Calgary, Canada, as a visiting scholar and in 2011 he spent 4 months at the Indian Institute of Science (IISc), Bangalore, India, as a visiting scholar. He is currently a Researcher at KTH coordinating the KTH Insurance Telematics Lab. He was a recipient of a Best Survey Paper Award by the IEEE Intelligent Transportation Systems Society in 2013.
\end{IEEEbiography}

\begin{IEEEbiography}[{\includegraphics[width=1in,height=1.25in,clip,keepaspectratio]{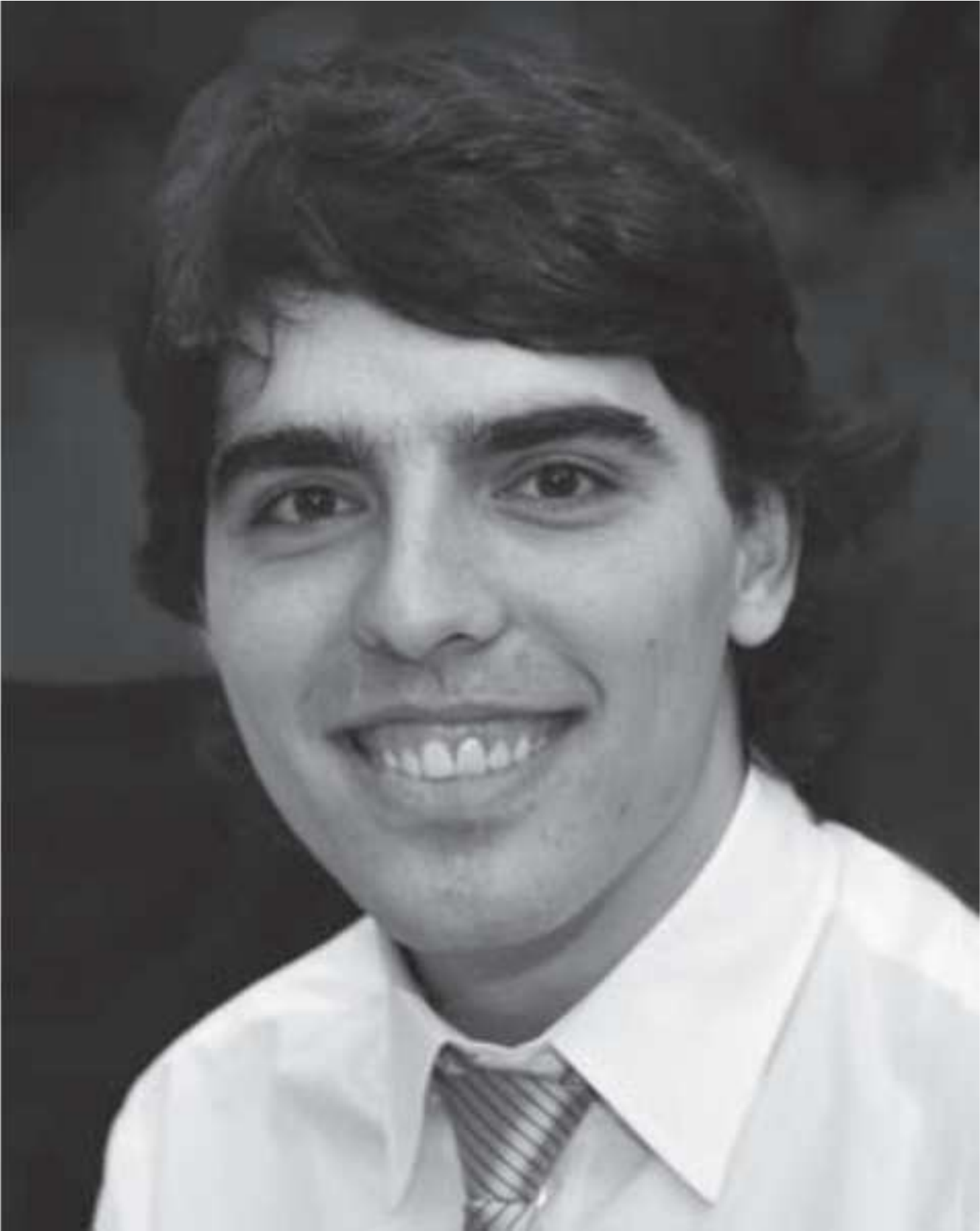}}]
{João G. P. Rodrigues}(S'11) received the MSc degree in electrical and computer engineering from the University of Porto, Porto, Portugal, in 2009. He is currently working toward the PhD degree with the University of Porto. He develops his work at the Institute for Telecommunications, and the main topics of his thesis are data gathering and mining in intelligent transportation systems. His main research interests include sensor networks and intelligent transportation systems. He received a Doctoral Scholarship from the Portuguese Foundation for Science and Technology in 2009.
\end{IEEEbiography}

\begin{IEEEbiography}[{\includegraphics[width=1in,height=1.25in,clip,keepaspectratio]{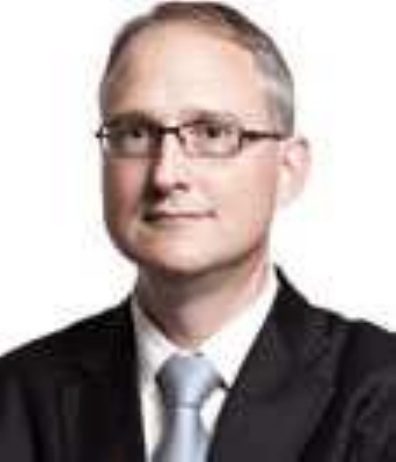}}]
{Peter H\"{a}ndel}(S'88-M'94-SM'98) received a PhD degree from Uppsala University, Uppsala, Sweden, in 1993. From 1987 to 1993, he was with Uppsala University. From 1993 to 1997, he was with Ericsson AB, Kista, Sweden. From 1996 to 1997, he was a Visiting Scholar with the Tampere University of Technology, Tampere, Finland. Since 1997, he has been with the KTH Royal Institute of Technology, Stockholm, Sweden, where he is currently a Professor of Signal Processing and the Head of the Department of Signal Processing. From 2000 to 2006, he held an adjunct position at the Swedish Defence Research Agency. He has been a Guest Professor at the Indian Institute of Science (IISc), Bangalore, India, and at the University of G\"avle, Sweden. He is a co-founder of Movelo AB. Dr. H\"andel has served as an associate editor for the IEEE TRANSACTIONS ON SIGNAL PROCESSING. He was a recipient of a Best Survey Paper Award by the IEEE Intelligent Transportation Systems Society in 2013.
\end{IEEEbiography}

\begin{IEEEbiography}[{\includegraphics[width=1in,height=1.25in,clip,keepaspectratio]{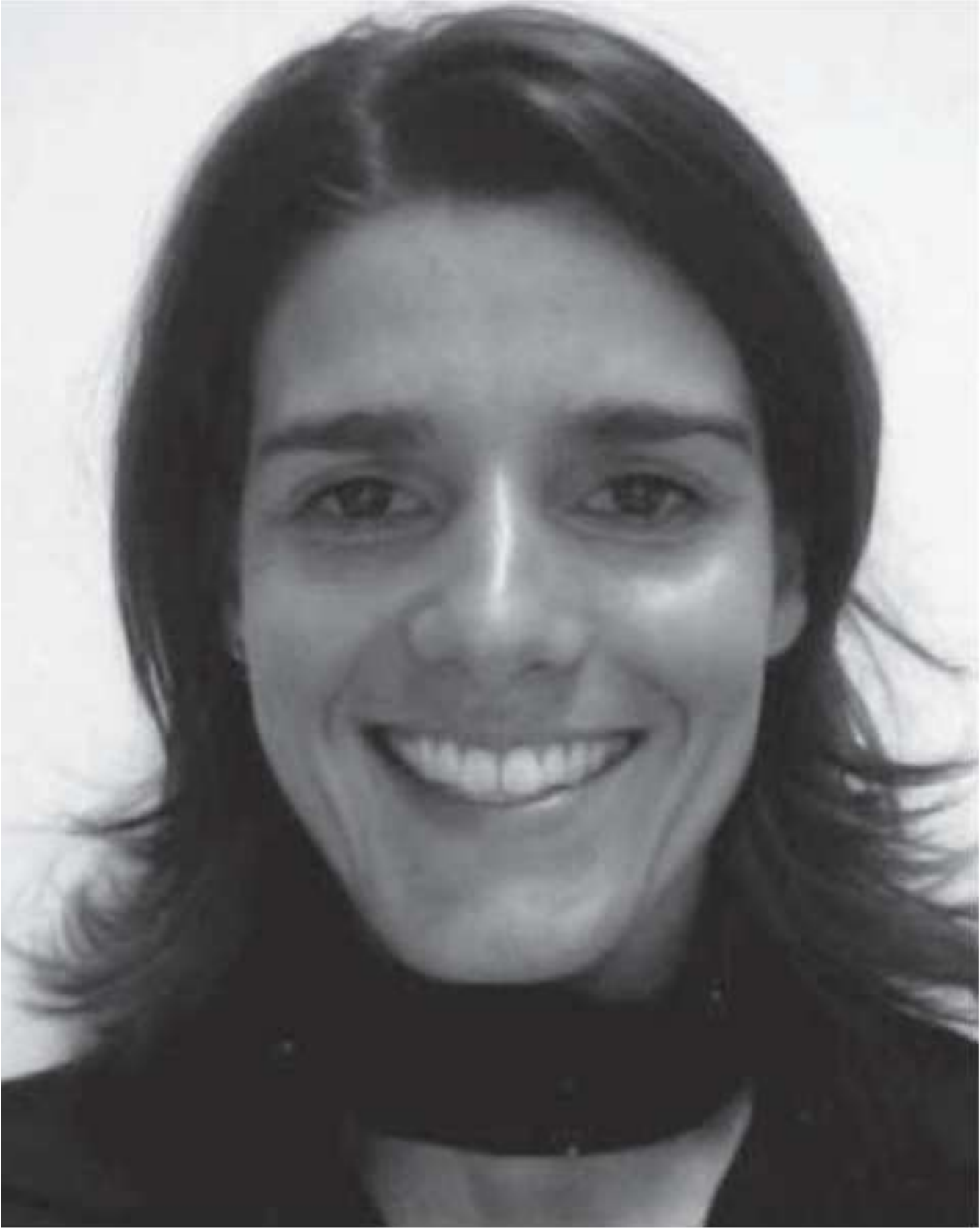}}]
{Ana Aguiar}(S'94-M'98-S'02-M'09) received the Electrical and Computer Engineering degree from the University of Porto, Porto, Portugal, in 1998, and the PhD in telecommunication networks from the
Technical University of Berlin, Berlin, Germany, in 2008. Since 2009, she has been an Assistant Professor
with the Faculty of Engineering, University of Porto. She began her career as an RF Engineer working for cellular operators, and she worked at Fraunhofer Portugal AICOS on service-oriented architectures and wireless technologies applied to ambient assisted living. She is the author of several papers published and presented in IEEE and ACM journals and conferences, respectively. She contributes to several interdisciplinary projects in the fields of intelligent transportation systems and well-being (stress). Her research interests include wireless networking and mobile sensing systems, specifically vehicular networks, crowd sensing, and machine-to-machine communications. She is a Reviewer for several IEEE and ACM conferences and journals.
\end{IEEEbiography}

% If you have an EPS/PDF photo (graphicx package needed) extra braces are
% needed around the contents of the optional argument to biography to prevent
% the LaTeX parser from getting confused when it sees the complicated
% \includegraphics command within an optional argument. (You could create
% your own custom macro containing the \includegraphics command to make things
% simpler here.)
%\begin{IEEEbiography}[{\includegraphics[width=1in,height=1.25in,clip,keepaspectratio]{mshell}}]{Michael Shell}
% or if you just want to reserve a space for a photo:

%\begin{IEEEbiography}{Michael Shell}
%Biography text here.
%\end{IEEEbiography}

% if you will not have a photo at all:
%\begin{IEEEbiographynophoto}{John Doe}
%Biography text here.
%\end{IEEEbiographynophoto}

% insert where needed to balance the two columns on the last page with
% biographies
%\newpage

%\begin{IEEEbiographynophoto}{Jane Doe}
%Biography text here.
%\end{IEEEbiographynophoto}

% You can push biographies down or up by placing
% a \vfill before or after them. The appropriate
% use of \vfill depends on what kind of text is
% on the last page and whether or not the columns
% are being equalized.

%\vfill

% Can be used to pull up biographies so that the bottom of the last one
% is flush with the other column.
%\enlargethispage{-5in}

% that's all folks
\end{document}